\documentclass[preprint,aps,nofootinbib,preprintnumbers,amsmath,amssymb,11pt]{revtex4}
\usepackage{hyperref}
\usepackage{amsthm}
\usepackage{graphicx}
\usepackage{amsfonts}
\usepackage[figuresright]{rotating}
\usepackage{amssymb}
\usepackage{amsmath}
\usepackage{psfrag}
\usepackage{subfigure}
\usepackage{multirow}
\usepackage{tabularx}

\usepackage{epsfig}
\usepackage{amsmath}
\usepackage{amssymb}
\usepackage{amsfonts}
\usepackage{graphicx}
\usepackage{dcolumn}
\usepackage{bm}
\usepackage{natbib}
\newcommand{\half}{\frac{1}{2}}
\newcommand{\beq}{\begin{equation}}
\newcommand{\eq}{\end{equation}}
\newcommand{\bea}{\begin{eqnarray}}
\newcommand{\ea}{\end{eqnarray}}
\newcommand{\p}{\partial}
\newcommand{\nn}{\nonumber}

\def\avg#1{\langle#1\rangle}

\def\nn{\nonumber}

\begin{document}

\begin{flushright}
\hfill{SU-ITP-15/02}
\end{flushright}

\title{Holographic RG flows with nematic IR phases}
\author{\vspace*{1 cm}\large
Sera Cremonini$^{\,1}$, Xi Dong$^{\,2}$, Junchen Rong$^{\,1}$, Kai Sun$^{\,3}$}
\email{sera@physics.tamu.edu, xidong@stanford.edu, jasonrong@physics.tamu.edu, sunkai@umich.edu}
\affiliation{\vspace*{0.5 cm}$^1$ George and Cynthia Mitchell Institute for Fundamental Physics and Astronomy,
Texas A\&M University, College Station, TX 77843--4242, USA\\}
\affiliation{$^2$ Stanford Institute for Theoretical Physics, Department of Physics, Stanford University,
Stanford, CA 94305, USA\\}
\affiliation{$^3$ Department of Physics, University of Michigan, 450 Church Street, Ann Arbor, MI 48109, USA\vspace*{0.5 cm}}

\begin{abstract}
\vspace*{1 cm}
We construct zero-temperature geometries that interpolate between a Lifshitz fixed point in the UV and an IR phase
that breaks spatial rotations but preserves translations.
We work with a simple holographic model describing two massive gauge fields coupled to gravity and a neutral scalar.
Our construction can be used to describe RG flows in non-relativistic, strongly coupled quantum systems with nematic order in the IR.
In particular, when the dynamical critical exponent of the UV fixed point is $z=2$ and the IR scaling exponents are
chosen appropriately,
our model realizes holographically the scaling properties of the bosonic modes of the quadratic band crossing model.
\end{abstract}

\vspace*{3 cm}
\maketitle

\def\thesection{\arabic{section}}
\def\thesubsection{\arabic{section}.\arabic{subsection}}
\numberwithin{equation}{section}
\newpage

\section{Introduction}
\label{Introduction}

The past few years have seen increasing efforts to apply the tools of holography to probe
condensed matter systems that are strongly coupled and typically poorly understood.
The goal of this program is to provide qualitative as well as quantitative insights into the unconventional behavior
of such systems, and in particular into their dynamical properties, which are challenging to probe using traditional techniques.
As a result, we have seen the emergence of novel gravitational solutions
-- often encoding a number of broken symmetries --
which mirror the rich structure of quantum phases familiar to the condensed matter community.
Within this holographic program, geometries that break translations (and sometimes rotations) have received much attention
recently.
In fact, the breaking of translational invariance (as a way to incorporate lattice effects) has been recognized as
a crucial ingredient for achieving a more realistic description of the conductive behavior of strongly correlated
electron systems
(see \emph{e.g.} \cite{Hartnoll:2012rj,Horowitz:2012ky,Horowitz:2012gs,Donos:2012js,Vegh:2013sk,Donos:2013eha,Andrade:2013gsa,Donos:2014yya}).

On the other hand, a gravitational description of the nematic phase, in which spatial rotations are the broken symmetries, has been largely unexplored. 
In particular, the question of how to obtain such phases in the infrared (IR)
via renormalization group (RG) flow from a non-relativistic fixed point in the ultraviolet (UV) is still open.
In this paper we would like to use the tools of holography to describe quantum systems which exhibit Lifshitz scaling in the UV,
where they are rotationally and translationally invariant, and flow to a phase in the IR which breaks spatial rotations.
If the latter phase has orientational order but no directional order (\emph{i.e.}, opposite directions are equivalent),
it describes a system which is nematic.
The order parameter for a nematic phase is a
\emph{director}, a symmetric traceless tensor.
Nonetheless, a vector $v^a$ can also be used to describe a nematic,
provided that its two possible directions $v^a$ and $-v^a$ are identified.
We will engineer the RG flows we are after by considering a simple phenomenological model in which gravity is coupled to
two massive abelian gauge fields and a neutral scalar, with the latter settling to a constant at the endpoints of the flow.
Our model has a global $\mathbb{Z}_2$ symmetry and therefore respects the $180^\circ$ rotational invariance crucial for nematics.
However, upon developing a vacuum expectation value, the gauge field $\tilde{A}$ responsible for generating the spatial anisotropy in the IR
can spontaneously break the $\mathbb{Z}_2$ symmetry\footnote{We are grateful to Sean Hartnoll for bringing this point to our attention.} --
a simple reflection of the fact that it is a vector and not a director.
Thus, to ensure the absence of directional order, we will require $\tilde{A}$ to live on $\mathbb{R}^2/\mathbb{Z}_2$,
where $\mathbb{R}^2$ denotes the spatial directions $x,y$ of the QFT.
In turn this will restrict the order parameter to live on $U(1)/\mathbb{Z}_2$, as desired for a nematic phase.

The concept of nematic ordering, which preserves the translational symmetry but breaks spontaneously the rotational symmetry
down to two-fold, was originally developed in the study of finite-temperature thermal phase transitions
in classical liquid crystals~\cite{Chaikin1995}.
Later it was discovered that quantum phases and quantum phase transitions with the same symmetry breaking pattern
can also arise at zero temperature in strongly correlated electronic systems \cite{Kivelson1998,Fradkin2010,Xu2008,Hartnoll2014}.
These quantum nematic phases have been observed in a wide range of materials and are proposed to be the key to
understand important properties of systems including high temperature
superconductors~\cite{TRANQUADA1995,Ando2002,Kohsaka2007,Hinkov2008,Mesaros2011},
bilayer ruthenates~\cite{Borzi2007,Raghu2009}, Fe-based superconductors~\cite{Chuang2010}, two-dimensional
electron gases~\cite{Fradkin2000}, fractional quantum Hall systems~\cite{Xia2011}, and doped manganites~\cite{Tao2014}.
Therefore, it is a valuable avenue to explore within the context of holography.

The construction of our paper applies generically to $2+1$ dimensional QFTs which
exhibit the following Lifshitz symmetry in the UV,
\beq
\label{UVscaling}
t \rightarrow \lambda^z t \, , \qquad  x_i \rightarrow \lambda x_i \, ,
\eq
with $x_i=\{x,y\}$ denoting the two spatial dimensions,
while in the IR exhibit the scaling
\beq
\label{IRscaling}
t \rightarrow \lambda^q t \, , \qquad  x \rightarrow \lambda^p x \, , \qquad y \rightarrow \lambda^q y \, .
\eq
In holography, Lifshitz fixed points can be geometrized by using a metric of the form
\beq
\label{Lifshitzmetric}
ds^2_{UV} = - r^{2z} dt^2 + \frac{dr^2}{r^2} + r^2 \left( dx^2 + dy^2 \right) \, ,
\eq
which is invariant under (\ref{UVscaling}) as long as we scale $r \rightarrow \lambda^{-1} r$, and reduces to $AdS_4$ when $z=1$.
Similarly, (\ref{IRscaling}) can be realized geometrically by choosing
\beq
\label{IRmetric}
ds^2_{IR} = - r^{2q} dt^2 + \frac{dr^2}{r^2} + r^{2p} dx^2 + r^{2q} dy^2 \, ,
\eq
which is clearly invariant under (\ref{IRscaling}), again provided that $r \rightarrow \lambda^{-1} r$.
Note that it preserves translations along the boundary directions $\{t,x,y\}$, but breaks spatial rotations,
unlike (\ref{Lifshitzmetric}), due to the different scaling of the $x$ and $y$ coordinates.
By appropriately choosing the potential for the scalar field in our model,
we will construct RG flows connecting the UV fixed point (\ref{UVscaling}) to
IR phases described by (\ref{IRscaling}).
We will focus in particular on the values $z=2$, $p=2$, and $q=3$, and construct a domain-wall solution which interpolates between
(\ref{Lifshitzmetric}) and (\ref{IRmetric}).

While our setup is general enough to encompass flows to a variety of nematic phases, in this paper we would like to focus on
applying it to a specific setting, namely that of the \emph{quadratic band crossing model}.
Indeed, as we will describe in detail in Section \ref{QBCmodel}, the bosonic modes of the quadratic band crossing model
exhibit interesting and quite non-trivial scalings depending on the energy scale of the physics that is being probed.
At the UV fixed point, both bosonic and fermionic modes obey the Lifshitz symmetry (\ref{UVscaling}), with the value of the
dynamical critical exponent given by $z=2$.
In the IR, however, the (strong) interactions between the fermions and the bosons change the dynamics -- and the scalings -- of the latter.
In particular, after integrating out the effects of the fermionic degrees of freedom,
one finds that the bosonic modes scale according to (\ref{IRscaling}), with the specific exponents given by $p=2$ and $q=3$
according to a one-loop self-energy correction.
Thus, our holographic setup provides a first step towards describing the physics of the quadratic band crossing model,
although we emphasize that it has broader applicability to IR nematic phases more generally.
Finally, while in this paper we have used a massive vector field
to break spatial rotations in the IR (and required it to live on $\mathbb{R}^2/\mathbb{Z}_2$ to describe nematic order),
we may also make a more general choice for the matter content of the bulk theory so that the order parameter is a director.
We leave this to future work \cite{WorkInProgress}.

The outline of the paper is as follows.
Section \ref{QBCmodel} discusses the quadratic band crossing model and in particular the scaling of its bosonic modes.
Section \ref{Setup} contains our holographic model and the structure of perturbations about the IR and UV fixed points.
In Section \ref{Numerics} we construct numerically a domain-wall solution interpolating between
the two fixed points, focusing on $z=2$, $p=2$, and $q=3$.
We conclude in Section \ref{Conclusions} with final remarks.

\section{The quadratic band crossing model}
\label{QBCmodel}

For electrons in a solid, two different energy bands may share the same energy at certain momentum points, which are known as
band crossing points.
A famous example of this type is graphene, in which two bands touch each other at the $K$ and $K'$ point of the Brillouin zone.
Near these two band crossing points, the energies of the two bands scale linearly with the momentum,
$\omega\propto \pm|\bold{q}|$, where the momentum $\bold{q}$ is measured
from a band crossing point. There, the low-energy physics is
described by the Dirac theory with dynamical critical exponent $z=1$, and the band crossing point is referred to as a Dirac point.
In addition to Dirac points, band crossing points with higher values of $z$, \emph{e.g.} $z=2$, also exist. A 2+1 dimensional example can be
found in bilayer graphene, a stack of two graphene layers, which has been experimentally realized and studied
(see for example Refs.~\cite{Nilsson2008,CastroNeto2009,Kotov2012} and references therein).
Such a $z=2$ band crossing point is known as a quadratic
band crossing point.

A quadratic band crossing point can be described by a two-component fermion spinor, $\bold{\Psi}=(\psi_1,\psi_2)$.
In $2+1$ dimensions, the action takes the following form
\begin{align}
S=\int d\bold{r} dt \;\bar{\bold{\Psi}}[\gamma_0(i\partial_0+t_0 \nabla^2)+\gamma_1
t_1 (\partial_x^2-\partial_y^2)+\gamma_2 (2\partial_x\partial_y)]\bold{\Psi} \, ,
\end{align}
where $t_0$ and $t_1$ are two control parameters; $\bar{\bold{\Psi}}=\bold{\Psi}^\dagger \gamma_0$ and
\begin{align}
\gamma_0=
\left(\begin{matrix}
0&-i\\
i&0
\end{matrix}\right)
\;\;\;\;\;\;
\gamma_1=
\left(\begin{matrix}
0&i\\
i&0
\end{matrix}\right)
\;\;\;\;\;\;
\gamma_2=
\left(\begin{matrix}
-i&0\\
0&i
\end{matrix}\right)
\;\;\;\;\;\;
\gamma_3=
\left(\begin{matrix}
1&0\\
0&1
\end{matrix}\right)
\end{align}
are the $2+1$ dimensional $\gamma$-matrices. For band crossing points in a solid, $t_0$ and $t_1$ can take arbitrary values.
Here we assume that the system preserves the charge conjugation symmetry (particle-hole) and is
invariant under  SO(2) space rotations. These two symmetries uniquely fix the control parameters $t_0=0$ and $t_1=1$ and fully
determine the quadratic terms in the action
\begin{align}
S_{\textrm{fermion}}=\int d\bold{r}dt \;\bar{\bold{\Psi}}[i\partial_0 \gamma_0+\gamma_1
(\partial_x^2-\partial_y^2)+\gamma_2 (2\partial_x\partial_y)]\bold{\Psi} \, .
\end{align}
In the absence of interactions, it is easy to realize that this fermion has a quadratic dispersion relation
$\omega=\pm k^2$, indicating that $z=2$.\footnote{This system is a marginal case between a metal and an insulator. Although the fermionic modes are gapless in analogy to a metal, due to the zero fermion density there is no Fermi surface. As shown below, the absence of a Fermi surface greatly simplifies the RG analysis.}

To describe interactions, we couple this fermion to bosonic modes.
As shown in Ref.~\cite{Sun2009}, to  leading order this fermion can be coupled with four different bosonic modes
$\Phi_i$, $i=0$, $1$, $2$, and $3$,
\begin{align}
S_{\textrm{couplings}}=\sum_{i=0}^3 g_i \int d\bold{r}dt \; \Phi_i \bar{\bold{\Psi}} \gamma_i \bold{\Psi} \, ,
\end{align}
where the $g$'s are the coupling constants.
To ensure that the action is local, \emph{i.e.} that there is no long-range interaction between fermions,
we require the boson modes to be gapped. Without loss of generality, the gap can be set to unity
\begin{align}
S_{\textrm{bosons}}=-\sum_{i=0}^3 \int d\bold{r}dt \; \Phi_i^2 \, .
\label{eq:action_boson}
\end{align}
Here we ignore terms with derivatives in $S_{\textrm{bosons}}$, because they are less relevant in the IR.
For the action $S=S_{\textrm{fermion}}+S_{\textrm{bosons}}+S_{\textrm{couplings}}$, dimension counting indicates
that the system is scaling invariant with $z=2$, \emph{i.e.} $[q_x]=[q_y]=1$, $[\omega]=2$, $[\bold{\Psi}]=1$, $[\Phi_i]=2$, and $[g_i]=0$.
The fact that $[g_i]=0$ implies that the couplings are marginal at tree level.
To go beyond tree level, we first integrate out the bosons, which results in a four-Fermi interaction
\begin{align}
S=-g\int d\bold{r}dt \; \psi_1^\dagger\psi_2^\dagger\psi_2\psi_1 \, ,
\end{align}
with $g=(g_0^2-g_1^2-g_2^2-g_3^2)/2$.
A one-loop RG calculation has revealed that this interaction is
marginally irrelevant in the IR if $g<0$ (attractive) and marginally relevant for $g>0$ (repulsive) \cite{Sun2009}.

Here we focus on the $g>0$ regime. In the UV, because the coupling scales down to zero, the
scaling law is dictated by dimension counting and thus we find
$z=2$ (for both bosons and fermions).
In the IR, however, the interaction grows to infinity, and a phase transition will take place.
Among the four bosonic modes, $\Phi_0$ describes the charge fluctuations and the other three bosonic modes are
order parameters for various symmetry breaking phases. If $\Phi_3$ obtains a nonzero expectation value,
the fermion gets a finite mass and becomes gapped, $\omega=\pm \sqrt{\avg{\Phi_3}^2+k^4}$.
This phase breaks spontaneously the time-reversal symmetry and is a topologically nontrivial quantum Hall insulator.
The other two modes, $\Phi_1$ and $\Phi_2$, form an $l_z=2$ representation of the rotational group $SO(2)$. They
give the order parameter of a nematic phase. If $\sqrt{\avg{\Phi_1}^2+\avg{\Phi_2}^2}$ is nonzero,
the $SO(2)$ rotational symmetry is broken spontaneously down to $Z_2$, \emph{i.e.} only the $180^\circ$ rotation
remains a symmetry operation. For $g>0$, whether the time-reversal or
the rotational symmetry will be broken in the IR is determined by microscopic details.
Mean-field analysis for some lattice models indicates that the gapped topological insulator phase is favored at small $g$,
while strong coupling (large $g$) favors the nematic phase~\cite{Sun2009}.

To better understand the nematic phase, without loss of generality we examine a case with $\avg{\Phi_1}>0$ and $\avg{\Phi_2}=0$.
It is easy to realize that here the massless Goldstone fluctuation is described by $\Phi_2$.
To quadratic order, the action of the fermions becomes
\begin{align}
S_{\textrm{fermion}}=\int d\bold{r}dt \;\bar{\bold{\Psi}}[i\partial_0 \gamma_0+\gamma_1
(\partial_x^2-\partial_y^2+\avg{\Phi_1})+\gamma_2 (2\partial_x\partial_y)]\bold{\Psi} \, .
\end{align}
This action contains two Weyl fermion points at momenta $\bold{k}_+=\left(0,\sqrt{\avg{\Phi_1}}\right)$ and
$\bold{k}_-=\left(0,-\sqrt{\avg{\Phi_1}}\right)$. If we go to momentum space and expand the action near
either of these two momentum points, a Weyl fermion action is obtained
\begin{align}
S_{\textrm{fermion}}=\int d\bold{q}\, d\omega \;\bar{\bold{\Psi}}[\gamma_0\omega\pm c (\gamma_1 q_y-\gamma_2 q_x) ]
\bold{\Psi}+O(q_x^2,q_y^2) \, ,
\end{align}
where $q_x$ and $q_y$ are measured from either of the two Weyl points $\bold{k}_+$ or $\bold{k}_-$, and the speed of light $c$
 is determined by the order parameter $c=2\sqrt{\avg{\Phi_1}^2+\avg{\Phi_2}^2}$.
In summary, in the nematic phase the quadratic band crossing point at $\bold{k}=\bold{0}$ splits into two Weyl points
at $\bold{k}_+$ or $\bold{k}_-$, and the direction along which they split is determined by spontaneous symmetry breaking
(for the case that we considered here, the splitting is along the $y$ axis).
Because these two Weyl points are related by space inversion, they combine together and form a Dirac fermion with $z=1$,
in analogy to the Dirac fermion in graphene.
In this nematic phase, the fermionic mode has $z=1$ at low energy and the reduction from $z=2$ in the UV to $z=1$ in the IR cures
the instability.

Although the fermionic mode has $z=1$, the low-energy boson modes in the nematic phase show different and interesting
scaling relations. Here, we write down a phenomenological action for the Goldstone mode $\Phi_2$
by including all terms allowed by symmetry. To leading order,
the action takes the form
\begin{align}
S_{\Phi_2}=\int  d\bold{r}dt \;  [\alpha_0 (\partial_t \Phi_2)^2-\alpha_1 (\partial_x \Phi_2)^2-\alpha_2 (\partial_y^2\Phi_2)] \, ,
\label{eq:action_Phi_2}
\end{align}
where the $\alpha$'s are three control parameters. A key difference between this action and the one shown in Eq.~\eqref{eq:action_boson}
lies in the fact that here $\Phi_2$ is gapless. This is because it is a Goldstone mode in the nematic phase. In the absence of the mass term,
terms with derivatives become important  and can no longer be ignored.
Because $\Phi_2$ is coupled to the fermionic modes $\bold{\Psi}$,
the fermions will introduce self-energy corrections and modify the dynamics of the $\Phi_2$ mode as we integrate out the fermionic degrees
of freedom.
At low energies, we can treat the fermionic mode as a Dirac mode as shown above.
Within this approximation, the one-loop self-energy correction is~\cite{Pisarski1984,Appelquist1986,Gonzalez1994595,Kotov2012}
\begin{align}
\Pi^{(2,2)}(\bold{q},\omega)=\frac{1}{8 c^2}\frac{c^2 q_y^2 - \omega^2}{\sqrt{c^2 q_x^2+c^2 q_y^2-\omega^2}} \, ,
\end{align}
and thus the effective theory for the $\Phi_2$ mode is
\begin{align}
S_{\textrm{eff}}=\int  d\bold{q} \, d\omega\; \left(\alpha_0 \, \omega^2-\alpha_1 q_x^2-\alpha_2 q_y^2
-\frac{1}{8 c^2}\frac{c^2 q_y^2 - \omega^2}{\sqrt{c^2 q_x^2+c^2 q_y^2-\omega^2}}\right)|\Phi_2(\bold{q},\omega)|^2 \, .
\end{align}
In the IR, the self-energy correction dominates over the $\alpha_0 \, \omega^2$ and $\alpha_2 \, q_y^2$ terms in the original action and thus significantly changes the scaling behavior of the Goldstone mode. In the limit $q_x\gg q_y$, $\omega$, we have
$\alpha_0 \omega^2-\alpha_1 q_x^2-\alpha_2 q_y^2\sim-\alpha_1 q_x^2$ and $c^2 q_x^2+c^2 q_y^2-\omega^2\sim c^2 q_x^2$,
and thus the effective action becomes
\begin{align}
S_{\textrm{eff}}=\int  d\bold{q} \, d\omega\; \left(
\frac{1}{8 c^3}\frac{\omega^2-c^2 q_y^2}{|q_x|}-\alpha_1 q_x^2\right)|\Phi_2(\bold{q},\omega)|^2 \, .
\end{align}
Note that this action is scale invariant under
\beq
\label{QBCscalings}
q_x\rightarrow \lambda^2 q_x \, , \qquad q_y\rightarrow \lambda^3 q_y \qquad \text{and} \quad
\omega\rightarrow \lambda^3 \omega \, ,
\eq
if we also scale the field $\Phi_2$ accordingly.

It is worthwhile to emphasize that although the above scaling law is based on a one-loop self-energy calculation, the structure of the
scaling relation may be more universal. For a Dirac system with $SO(2,1)$ symmetry, the self-energy correction $\Pi^{(2,2)}$
must be invariant under $SO(1,1)$ rotations, i.e. Lorentz boost for $t$ and $y$, and thus $\omega$ and $q_y$ are expected to have the
same scaling behavior. Therefore, in the strong coupling limit, one may expect the scaling relation
\beq
q_x\rightarrow \lambda^p q_x \, , \qquad q_y\rightarrow \lambda^q q_y \qquad \text{and} \quad \omega\rightarrow \lambda^q \omega \, ,
\eq
although the values of $p$ and $q$ may deviate from the one-loop result.
Next, we will examine how such scalings can be realized geometrically, using holography.

\section{The holographic model}
\label{Setup}

We are interested in building a model which will admit a zero-temperature flow from a non-relativistic
UV fixed point described by (\ref{UVscaling}) to an IR fixed point in which spatial rotations are broken,
described by (\ref{IRscaling}).
Geometrically, this entails finding domain-wall solutions that connect (\ref{Lifshitzmetric}) in the UV
to (\ref{IRmetric}) in the IR.
Before introducing the particular model we will be working with, we would like to remind the reader that
both (\ref{Lifshitzmetric}) and (\ref{IRmetric}) are \emph{exact solutions} to the theory of
a massive abelian $U(1)$ gauge field coupled to gravity \cite{Kachru:2008yh,Taylor:2008tg},
\beq
\label{action}
S = \int d^4 x \sqrt{-g} \left( R - \frac{1}{4} F^2 - \half W A^2 - V \right) \, ,
\eq
where $W$ and $V$ are constant.
In particular, the Lifshitz metric (\ref{Lifshitzmetric}) can be supported by (\ref{action})
when the gauge field is purely electric and is given by
\beq
\label{gaugeUV}
A_\mu = (A_1 \, r^z, \, 0,0,0) \, ,
\eq
with $A_1^2 = 2(z-1)/z \, , W = 2z$ and $V = -(4+z+z^2) $.
On the other hand, the IR metric (\ref{IRmetric}) is an exact solution to (\ref{action})
when the gauge field is oriented along the $x$ direction,
\beq
\label{gaugeIRgen}
A_\mu = (0,0,A_2 \, r^p, \, 0) \, ,
\eq
and the remaining parameters are related via
$ A_2^2 = 2(q-p)/p\, , W = 2 \, p \, q$ and $V = -(4 q^2 + p^2 + pq) $.
Domain-wall solutions interpolating between different Lifshitz geometries, or between Lifshitz and AdS,
have been constructed in a number of setups, typically involving gravity coupled to a massive $U(1)$ gauge field
-- and sometimes a scalar -- and their generalizations (see e.g. \cite{Kachru:2008yh,Braviner:2011kz,Liu:2012wf,Kachru:2013voa}).
As a concrete example, flows of this type can be obtained (see e.g. \cite{Liu:2012wf})
in simple phenomenological models of the form
\beq
\label{L1}
\mathcal{L} = R -\half (\p \phi)^2 - \frac{1}{4}F^2 - V(\phi) - W(\phi) A^2 \, ,
\eq
in the presence of a background electric field $A = A_t(r) dt$.

Here, however, we are also interested in breaking spatial rotations in the IR, where we want the geometry to be described by
(\ref{IRmetric}), with different spatial directions exhibiting distinct scalings.
To this end, a simple way to generalize the model (\ref{L1}) is to add a second massive gauge field $\tilde{A}$.
Thus, we take our working model to be of the form
\beq
\mathcal{L} = R -\half (\p \phi)^2 - \frac{1}{4}Z F^2 - \half W(\phi) A^2
- \frac{1}{4}\tilde{Z}\tilde{F}^2  - \half \tilde{W}(\phi) \tilde{A}^2 - V(\phi) \, ,
\label{ourmodel}
\eq
with  $ F = dA$ and $ \tilde{F}=d\tilde{A}$, and
the gauge fields $A$ and $\tilde{A}$ chosen so that
\beq
\label{gaugeansatz}
A_\mu = (A_t(r) \, ,0, 0,  0) \, , \qquad \tilde{A}_\mu = (0,0, A_x(r), \, 0) \, .
\eq
They will dominate the geometry in the UV and IR, respectively, where they will become $A_t^{UV} \propto r^z$ and $A_x^{IR} \propto r^p$,
generating a Lifshitz solution at high energies and one 
with broken spatial rotations in the opposite, low-energy regime.
The scalar potential $V$ (or more precisely, the effective scalar potential
$V_{eff}(\phi) = V(\phi) + \half W(\phi) A^2 + \half \tilde{W}(\phi) \tilde{A}^2$)
will be chosen to drive the scalar field to constant values in the IR and UV, $\phi = \{\phi_{IR},\phi_{UV}\}$.
Finally, we will be interested in solutions described by a diagonal metric,
\beq
\label{metricansatz}
ds^2 = - f(r) dt^2 + \frac{dr^2}{r^2} + g_1(r) dx^2 + g_2 (r) dy^2 \, ,
\eq
where the two functions $g_1(r)$ and $g_2(r)$ will not generically be the same, in order to allow for the breaking of rotational
symmetry at low energies.

At this stage we would like to elaborate on the role of the vector field $\tilde{A}_\mu$
responsible for the absence of spatial rotational symmetry in the IR.
As we discussed briefly in the Introduction, a nematic phase has orientational order but no directional order -- opposite
directions in the $x, y$ plane are completely equivalent.
As a consequence, its order parameter must respect the discrete $\mathbb{Z}_2$ symmetry (in particular, it must live on
$U(1)/\mathbb{Z}_2$ for the case of two spatial dimensions) and is generically described by a symmetric traceless tensor, \emph{i.e.} a director.
While a vector can still be used to describe nematic ordering, in order to do so its direction must be identified with the opposite one.

For our model, this means that the abelian gauge field $\tilde{A}_\mu$
must live on $\mathbb{R}^2/\mathbb{Z}_2$ rather than simply $\mathbb{R}^2$.
One possible way to realize the latter is by gauging the global $ \mathbb{Z}_2$ symmetry of our theory
(see \emph{e.g.} \cite{Krauss:1988zc,Banks:2010zn}).
We expect that there shouldn't be obstructions to doing so in the gravity theory, at least from an effective low-energy point of view.
Whether this model can be UV completed is of course an important question which we don't attempt to address here. In fact,
it would be more desirable to choose the matter content of the model to reflect more directly the presence of a director, without resorting
to using a vector.
Nonetheless, the operation of gauging the global $ \mathbb{Z}_2$ would then ensure that the overall sign of the gauge field is not physically observable,
and that the order parameter of the IR phase is essentially $\tilde{A}^2$.
The trick of using the square of a vector to describe nematic order
may work best in two spatial dimensions, which is the case we considered here, but fail or be more subtle in higher dimensions.
This is because in two spatial dimensions a director and a vector have the same number of degrees of freedom, but this is no longer true in higher
dimensions.

Finally, we should note that a spatial vector can be used as the order parameter for a ferromagnetic or a ferroelectric phase.
In addition to rotational symmetry breaking,  ferromagnetic order also breaks the time-reversal symmetry
(the magnetization changes sign under time-reversal symmetry), while nematic or ferroelectric order is invariant under time-reversal.
For our gravity setup, if we require the IR phase to preserve the time-reversal symmetry,  
the gauge field should be restricted to live on $\mathbb{R}^4/\mathbb{Z}_2$, which implies that our order parameter is in fact a 
director, instead of a vector, and thus the IR phase is nematic.

With these ingredients in place we will be able to engineer
the flow we are after, in which the rotational symmetry is restored in the UV but the system is non-relativistic at every energy
scale.
While there should be alternative ways to build such systems, this particular model will allow us to work with simple ordinary differential equations,
and to easily decouple some of the IR and UV perturbations from the rest, thus simplifying the analysis significantly.
Also, at this stage our model is entirely phenomenological. Whether it can be embedded into string theory
constructions is an interesting question, but is beyond the scope of our paper.
We leave a more extensive analysis of how to engineer
IR nematic phases to future work.

\subsection{Domain-wall solutions}

The equations of motion for our model (\ref{ourmodel}) are given by
\bea
\label{eom}
&& R_{\mu\nu} = \half \p_\mu \phi \p_\nu \phi + \frac{W}{2}A_\mu A_\nu + \frac{\tilde{W}}{2}\tilde{A}_\mu \tilde{A}_\nu
+ \frac{Z}{2} F_{\mu\rho} F_\nu^{\;\;\rho} + \frac{\tilde{Z}}{2} \tilde{F}_{\mu\rho} \tilde{F}_\nu^{\;\;\rho}
+ \frac{g_{\mu\nu}}{8} \left[4V - Z F^2 - \tilde{Z} \tilde{F}^2 \right] \, , \nn \\
&&\frac{1}{\sqrt{-g}} \p_\mu \left( \sqrt{-g} \p^\mu \phi \right) =
\half W^{\prime} A^2
+ \half \tilde{W}^{\, \prime} \tilde{A}^2 +  V^\prime \, , \nn \\
&& \frac{1}{\sqrt{-g}} \p_\mu \left( \sqrt{-g} Z F^{\mu \nu}  \right) = W A^\nu \, , \nn \\
&& \frac{1}{\sqrt{-g}} \p_\mu \left( \sqrt{-g} \tilde{Z} \tilde{F}^{\mu \nu}  \right) = \tilde{W} \tilde{A}^\nu \, ,
\ea
where primes denote derivatives with respect to the scalar field.
Recall that our metric ansatz is (\ref{metricansatz}) and we assumed that the gauge fields have the simple form
given in (\ref{gaugeansatz}).

In the IR the geometry is described by the following background solution,
\bea
\label{IRbackground}
ds^2 &=& - r^{2q} dt^2 + \frac{dr^2}{r^2} + r^{2p} dx^2 + r^{2q} dy^2 \, , \nn \\
A_\mu &=& (0,0, 0, 0) \, , \qquad  \tilde{A}_\mu = (0,0, r^p, 0) \, , \qquad \phi = \phi_{IR}\, ,
\ea
provided that the scalar potential $V(\phi)$ and coupling $\tilde{W}(\phi)$ obey
\bea
&& V^\prime (\phi_{IR}) + \half \tilde{W}^{\, \prime} (\phi_{IR})=0 \, , \qquad  V(\phi_{IR}) = -4q^2 -p^2 -pq \, , \\
&& \tilde{W} (\phi_{IR}) = 4 q (q-p) \, , \qquad \tilde{Z} =\frac{2(q-p)}{p} \, .
\ea
Note that since $A_t$ vanishes to leading order in the IR geometry, the background equations of motion do not constrain the value of $W(\phi_{IR})$.
For the purpose of describing the quadratic band crossing model we discussed in Section \ref{QBCmodel},
we will eventually set $p=2$ and $q=3$. However, for now we will keep these scalings arbitrary.

In the UV, on the other hand, the background is given by
\bea
\label{UVbackground}
ds^2 &=& - r^{2z} dt^2 + \frac{dr^2}{r^2} + r^{2} dx^2 + r^{2} dy^2 \, , \nn \\
A_\mu &=& (r^z,0, 0, 0) \, ,  \qquad \tilde{A}_\mu = (0,0,0, 0)\, , \qquad  \phi = \phi_{UV}\, ,
\ea
supported by the following
\bea
&& V^\prime (\phi_{UV}) - \half W^\prime(\phi_{UV}) =0 \, , \qquad  V(\phi_{UV})=-z^2 -z -4 \, ,\\
&& W(\phi_{UV}) = 4(z-1) \, , \qquad Z = \frac{2(z-1)}{z} \, .
\ea
It is now  $\tilde{W}(\phi_{UV})$ which is not constrained by the background equations of motion, again
because $A_x$ vanishes to leading order in the UV.
To satisfy the scaling of the quadratic band crossing model we will eventually need to take $z=2$.
For generality we will keep $z$ arbitrary for now.
We will also assume that the UV (IR) fixed point is a local maximum (minimum) of the effective potential, \emph{i.e.}
\beq
V_{eff}^{\prime\prime}(\phi_{UV}) < 0 \, , \qquad \text{and} \qquad V_{eff}^{\prime\prime}(\phi_{IR})>0 \, ,
\eq
where $V_{eff}(\phi) = V(\phi) + \half A^2 W (\phi) + \half \tilde{A}^2 \tilde{W}(\phi)$.
The scalar field will then roll from the maximum at $\phi=\phi_{UV}$ to the minimum at $\phi = \phi_{IR}$.
From the first equation in \eqref{eom} it is evident that our domain-wall solutions will satisfy the null energy
condition as long as $W$ and $\tilde W$ stay non-negative during the entire flow, as is true in the case that
will be numerically studied in Section \ref{Numerics}.

Next, we would like to study the response of the IR and UV geometries to linearized fluctuations, to ensure that we can
find a well-behaved RG flow between the two fixed points.

\subsubsection{IR perturbations}

We start by examining the equation of motion for the scalar field,
\beq
\label{scalarEOM}
\Box \phi = \frac{\p V}{\p \phi} +  \half A^2 \frac{\p W}{\p \phi} +  \half \tilde{A}^2 \frac{\p \tilde{W}}{\p \phi} \, .
\eq
Expanding the field in perturbations $\phi(r) = \phi_{IR} + \delta\phi^{(IR)}$ about its IR value and linearizing (\ref{scalarEOM})
we find that $\delta\phi^{IR}(r)$ obeys
\beq
\label{scalarpertEOM}
\left( r^2\p_r^2 + r (1+p+2q) \p_r - V^{\prime\prime} (\phi_{IR}) - \half \tilde{W}^{\prime\prime} (\phi_{IR}) \right) \delta\phi^{IR}(r)=
\tilde{W}^{\, \prime} (\phi_{IR}) \left( \half \delta g_{xx}^{IR} - \delta a_x^{IR} \right) \, .
\eq
Notice that the scalar perturbation decouples from the metric and gauge field fluctuations $\delta g_{xx}^{IR}$ and $\delta a_x^{IR}$ only when
$\tilde{W}^{\, \prime} (\phi_{IR}) =0$.
Thus, to simplify the analysis and ensure decoupling we impose
\beq
\tilde{W}^{\, \prime} (\phi_{IR}) =0 \quad  \Rightarrow \quad V^{\, \prime} (\phi_{IR}) =0 \, ,
\eq
where the second condition is needed to satisfy the background equations of motion.
We emphasize that this restriction on $\tilde{W}^{\, \prime} (\phi_{IR})$  is purely for convenience and is \emph{not} required
for a solution to exist.
Finally, solving (\ref{scalarpertEOM}) we have
\beq
\delta\phi^{(IR)} = \phi_\pm^{(IR)} \, r^{-\half p - q  \pm \half \sqrt{(p+2q)^2 + 4 V^{\prime\prime}(\phi_{IR})
+ 2 \tilde{W}^{\prime\prime}(\phi_{IR}) } }  \, .
\eq
When the scaling exponents are given by $p=2$ and $q=3$ the two modes reduce to
\beq
\delta\phi^{(IR)} = \phi_\pm \, r^{-4 \pm \sqrt{16 + V^{\prime\prime}(\phi_{IR}) + \half \tilde{W}^{\prime\prime}(\phi_{IR}) } }  \, .
\eq
To avoid turning on perturbations which are relevant about the IR fixed point, and which would therefore destabilize it,
when constructing RG flow we will choose
the mode which vanishes as $r\rightarrow 0$ and increases towards the boundary, as $r \rightarrow \infty$.
The perturbation corresponding to the negative root $\phi_-^{(IR)}$ is always divergent as the IR is approached, and thus needs to be turned off.
To ensure that the other perturbation approaches zero as $r\rightarrow 0$ we need to impose
\beq
V^{\prime\prime}(\phi_{IR}) + \half \tilde{W}^{\prime\prime}(\phi_{IR}) \geq 0 \, ,
\eq
which coincides with the requirement that $\phi = \phi_{IR}$ is a minimum of the effective scalar potential.

Next, we want to examine the behavior of the gauge field $A_t = 0 + \delta a_t^{(IR)}(r)$.
Since it vanishes to leading order in the IR, its perturbation also decouples from the other fluctuations, and obeys the equation
\beq
\left( r^2 \p_r^2 + r (1+p) \p_r - \frac{W(\phi_{IR})}{Z} \right) \delta a_t^{(IR)}(r) = 0 \, ,
\eq
with solution
\beq
\delta a_t^{(IR)}(r)  = a^{(IR)}_\pm r^{- \frac{p}{2} \pm \half \sqrt{p^2 + 4 W(\phi_{IR})/Z}} \, .
\eq
When $p=2$ this reduces to
\beq
\delta a_t^{(IR)}(r) = a_\pm r^{-1 \pm \sqrt{1+ W(\phi_{IR})/Z}} \, .
\eq
Thus, the scaling behavior of $A_t$ near the origin is controlled by the parameter $W(\phi_{IR})$, which is \emph{not} specified by the background
equations of motion. Again, the perturbation corresponding to the negative root is always divergent in the IR.
The positive root corresponds to an irrelevant perturbation (growing towards the UV) provided that $ W(\phi_{IR})/Z >0$.

The remaining perturbations, involving the metric and the $A_x$ gauge field, are all coupled to each other.
We will assume that they have an IR expansion of the form
\bea
g_{tt} &=& r^{2q} \left(1+\delta g_{tt}^{(IR)}\right) \, , \quad \text{with} \quad \delta g_{tt}^{(IR)} = \alpha_1 \, r^ \alpha \, , \\
g_{xx} &=& r^{2p} \left(1+\delta g_{xx}^{(IR)}\right) \, , \quad \text{with} \quad \delta g_{xx}^{(IR)} = \alpha_2 \, r^ \alpha \, , \\
g_{yy} &=& r^{2q} \left(1+\delta g_{yy}^{(IR)} \right) \, , \quad \text{with} \quad \delta g_{yy}^{(IR)} = \alpha_3 \, r^ \alpha \, , \\
\tilde{A}_x^{(IR)} &\sim& r^p \left(1+ \delta a_x^{(IR)}\right)  \quad \text{with} \quad \delta a_x^{(IR)} = \alpha_4 \,
r^\alpha \, ,
\ea
where we have left $g_{rr}=1/r^2$ untouched.

Solving for the scaling exponent $\alpha$ and the $\alpha_i$ coefficients  we find the following:
\begin{enumerate}
\item
a constant mode ($\alpha=0$) with:
\beq
\alpha_2 = 2\alpha_4  \quad \text{and $\alpha_1,\alpha_3,\alpha_4$ free} \,  ;
\eq
\item
a mode which scales with $\alpha=-p-2q$ and has
\beq
\alpha_1 = \alpha_2-\alpha_3 \, , \quad \alpha_4 = \frac{p^2+pq+4q^2}{4q(q-p)}  \alpha_2   \quad \text{and $\alpha_2,\alpha_3$ free} \, .
\eq
This mode always diverges as the IR is approached (it is relevant) under the assumption that $p$ and $q$ are positive,
and therefore needs to be discarded.
\item
a mode which scales with $\alpha = - \half p - q + \half \sqrt{9p^2+20 q^2 - 20 p q} $ and has
\bea
\alpha_1 &=& \alpha_3
\, , \qquad \alpha_2 = \frac{5p-2q + \sqrt{9p^2+20 q^2 - 20 p q}}{4q -3p - \sqrt{9p^2+20 q^2 - 20 p q} }   \, \alpha_3 \, \nn \\
\alpha_3 &=&  \left(\frac{q-p}{p}\right) \frac{-4q +3p + \sqrt{9p^2+20 q^2 - 20 p q}}{-5q +2p + \sqrt{9p^2+20 q^2 - 20 p q} }  \, \alpha_4
\quad \text{and $\alpha_4$ free}\, ;
\ea
This mode is irrelevant provided that $\sqrt{9p^2+20 q^2 - 20 p q} > p + 2q $.
\item
a mode which scales as $\alpha = - \half p - q - \half \sqrt{9p^2+20 q^2 - 20 p q} $ and has
\bea
\alpha_1 &=& \alpha_3
\, , \qquad \alpha_2 = \frac{5p-2q - \sqrt{9p^2+20 q^2 - 20 p q}}{4q -3p + \sqrt{9p^2+20 q^2 - 20 p q} }   \, \alpha_3 \, \nn \\
\alpha_3 &=&  \left(\frac{q-p}{p}\right) \frac{4q -3p + \sqrt{9p^2+20 q^2 - 20 p q}}{5q -2p + \sqrt{9p^2+20 q^2 - 20 p q} }  \, \alpha_4
\quad \text{and $\alpha_4$ free}\,.
\ea
This mode is also always relevant (assuming $p,q>0$) and thus must be discarded.
\end{enumerate}
Of these, the only mode which is irrelevant when the scaling exponents are $p=2$ and $q=3$ is the third one, which scales
as $\alpha = 2 \, (-2 + \sqrt{6}) $ and has:
\bea
\label{IRirrelevant}
\alpha_1 &=& \alpha_3 \, , \quad \alpha_2 = - \frac{2}{3} (\sqrt{6} + 3)\, \alpha_3 \,
\quad \alpha_4 = -\frac{1}{3}(2\sqrt{6}-3) \, \alpha_3 \, ,  \quad \text{with $\alpha_3$ free}\, .
\ea
Finally, counting all parameters we find 11 integration constants in the IR, 5 of which are associated with modes which are relevant
and therefore need to be set to zero when constructing RG flow, and 3 of which are associated with a constant mode.

\subsubsection{UV perturbations}

We can now examine the structure of the perturbations in the UV, for a generic value of the dynamical critical exponent $z$.
Expanding the scalar equation of motion (\ref{scalarEOM}) to linear order in perturbations, with
$\phi_{UV}(r) = \phi_{UV} + \delta\phi^{(UV)}(r)$, we find that the fluctuation obeys
\beq
\label{scalarpertEOMUV}
\left( r^2\p_r^2 + r (3+z) \p_r
- V^{\prime\prime} (\phi_{UV}) + \half W^{\prime\prime} (\phi_{UV}) \right) \delta\phi^{(UV)}(r)=
W^{\, \prime} (\phi_{UV}) \left( \half \delta g_{tt}^{(UV)} - \delta a_t^{(UV)} \right) \, .
\eq
As before, the scalar field fluctuation decouples from the other ones provided that $W^\prime (\phi_{UV}) =0$.
Again, we will assume that this is the case in order to simplify the analysis, although it is by no means a necessary condition.
In order to satisfy the background equations of motion we are then forced to take
$V^\prime (\phi_{UV}) =0$.
Solving (\ref{scalarpertEOMUV}) we find the following modes,
\beq
\delta\phi^{(UV)} = \phi^{(UV)}_\pm \, r^{-1-\frac{z}{2} \pm \half
\sqrt{(z+2)^2 +4 V^{\prime\prime}(\phi_{UV}) -2 W^{\prime\prime}(\phi_{UV}) } }
\, .
\eq
When $z=2$ they reduce to:
\beq
\delta\phi^{(UV)} = \phi^{(UV)}_\pm \, r^{-2 \pm \sqrt{4+V^{\prime\prime}(\phi_{UV}) - W^{\prime\prime}(\phi_{UV})/2 } } \, .
\eq
In the UV, it is the fluctuation of the $\tilde{A}$ gauge field which decouples from the other modes,
since the field vanishes to linear order.
Letting $ A_x =0 + \delta a^{(UV)}_x(r)$ we find
\beq
\left( r^2 \p_r^2 + r (z+1) \p_r - \frac{\tilde{W}(\phi_{UV})}{Z} \right) \delta a_x^{(UV)}(r) = 0 \, ,
\eq
whose solutions are given by
\beq
\delta a^{(UV)}_x = \tilde{a}_{\pm}^{(UV)} r^{-\frac{z}{2} \pm \half \sqrt{z^2 + \, 4 \, \frac{\tilde{W}(\phi_{UV})}{\tilde{Z}} } } \, .
\eq
Thus, the scaling of the perturbations
can be controlled by tuning the value of $\tilde{W}(\phi_{UV}) $, which is not determined by the background equations of motion.

The remaining fluctuations of the metric and the gauge field $A_t$ are all coupled together.
Parametrizing them in the following way
\bea
g_{tt} &=& r^{2z} \left(1+\delta g_{tt}^{(UV)}\right) \, , \quad \text{with} \quad \delta g_{tt}^{(UV)} = \beta_1 r^ \beta \, , \\
g_{xx} &=& r^{2} \left(1+\delta g_{xx}^{(UV)}\right) \, , \quad \text{with} \quad \delta g_{xx}^{(UV)} = \beta_2 r^ \beta \, , \\
g_{yy} &=& r^{2} \left(1+\delta g_{yy}^{(UV)} \right) \, , \quad \text{with} \quad \delta g_{yy}^{(UV)} = \beta_3 r^ \beta \, , \\
A_t^{(UV)} &=& r^z \left(1+ \delta a_t^{(UV)} \right)\, ,  \quad \text{with} \quad \delta a_t^{(UV)} = \beta_4 r^\beta \, ,
\ea
where we are leaving $g_{rr}=1/r^2$ untouched.
We find the following solutions:
\begin{enumerate}
\item
a constant mode ($\beta=0$) with
\beq
\label{mode1}
\beta_1 = 2 \beta_4 \quad \text{and $\beta_2,\beta_3,\beta_4$ free} \, ;
\eq
\item
a mode which scales with $\beta = - z -2 $ and has
\beq
\label{mode2}
\beta_1 = - \frac{4(z-1)}{4+z^2+z}\beta_4 \, , \quad \beta_3 = \beta_1-\beta_2 \quad \text{and $\beta_2, \beta_4$ free} \, ;
\eq
\item
a mode which scales with $\beta = -\frac{z}{2}-1 + \half\sqrt{9z^2-20z+20} $ and has
\bea
\beta_1 &=& \left[ \frac{3z}{2}-3 - \half\sqrt{9z^2-20z+20} \right] \beta_3 \, , \quad \beta_2 = \beta_3 \, , \nn \\
\beta_4 &=& \frac{z}{4(z-1)} \left( \sqrt{9z^2-20z+20} - 3z \right) \beta_3 \quad  \text{and $\beta_3$ free} \, ;
\label{mode3}
\ea
\item
a mode which scales with $\beta = -\frac{z}{2}-1 - \half\sqrt{9z^2-20z+20} $ and has
\bea
\beta_1 &=& \left[ \frac{3z}{2}-3 + \half\sqrt{9z^2-20z+20} \right] \beta_3 \, , \quad \beta_2 = \beta_3 \, , \nn \\
\beta_4 &=& -\frac{z}{4(z-1)} \left( \sqrt{9z^2-20z+20} + 3z \right) \beta_3 \quad  \text{and $\beta_3$ free} \, .
\label{mode4}
\ea
\end{enumerate}
Note that these perturbations were already found in \cite{Ross:2009ar}.
As in the IR, we have 11 integration constants, 3 of which come from the constant mode.

The case $z=2$, which is what we will focus on in Section \ref{Numerics}
to describe the quadratic band crossing model, needs to be
treated with some care \cite{Bertoldi:2009vn,Ross:2009ar,Cheng:2009df,Holsheimer:2013ula}.
Note that when $z=2$ the third mode above becomes a constant, while the second and fourth modes both scale as $r^{-4}$.
Indeed, for $z=2$ there are additional logarithmic modes which are not captured by our perturbation ansatz above and
which can change the asymptotic form of the geometry.
For a detailed discussion of these logarithmic modes
without relying on linearizing the equations of motion, we refer the reader to \cite{Cheng:2009df}.
Here however we will work with linearized equations.
The $z=2$ modes can then be taken into account
by writing down the following ansatz for the perturbations:
\bea
\delta g_{tt}^{(UV)} &=& (\beta_1 + \tilde\beta_1 \log r) r^ \beta \, , \\
\delta g_{xx}^{(UV)} &=& (\beta_2 + \tilde\beta_2 \log r) r^ \beta \, , \\
\delta g_{yy}^{(UV)} &=& (\beta_3 + \tilde\beta_3 \log r) r^ \beta \, , \\
\delta a_t^{(UV)} &=& (\beta_4 + \tilde\beta_4 \log r) r^ \beta \, .
\ea
We then find the following solutions, in agreement with the analysis of \cite{Ross:2009ar},
\begin{enumerate}
\item
$\beta=0$ with
\beq
\beta_1 = 2 \beta_4 -\tilde\beta_4 \,, \quad
\tilde\beta_1 = 2\tilde\beta_4 \,,\quad
\tilde\beta_2 = \tilde\beta_3 = -\tilde\beta_4 \quad
\text{and $\beta_2,\beta_3,\beta_4, \tilde\beta_4$ free} \,.
\eq
Since the leading logarithmic modes grow in the UV, we will impose $\tilde\beta_4=0$
as a boundary condition to ensure that the Lifshitz asymptotics are not changed \cite{Ross:2009ar}.
The role played by the leading log modes is actually quite rich. They were argued in \cite{Cheng:2009df}
to describe marginally relevant deformations of the theory.
Finally, note that when
$\tilde{\beta}_4=0$ this constant mode reduces to the one (\ref{mode1}) we had already identified for general $z$, which is a good consistency
check.
\item
$\beta = -4$ with
\begin{gather}
\beta_1 = -\frac{2}{5} \beta_4 + \frac{3}{5}\tilde\beta_4 \,, \quad
\beta_2 = -\frac{2}{5} \beta_4 -\beta_3 - \frac{2}{5}\tilde\beta_4 \,, \quad
\tilde\beta_1 = -\frac{2}{5} \tilde\beta_4 \,,\nonumber\\
\tilde\beta_2 = \tilde\beta_3 = -\frac{1}{5} \tilde\beta_4 \quad
\text{and $\beta_3,\beta_4, \tilde\beta_4$ free} \,. \label{-4log}
\end{gather}
Again, when $\tilde{\beta}_4=0$ we recover the mode $\sim r^{-4}$ which we had found in (\ref{mode2}) and (\ref{mode4}).
\end{enumerate}

As before, we find that there are 11 integration constants in the UV.
Here, we choose to turn off the leading log mode (i.e. setting $\tilde{\beta}_4=0$ in the $\beta=0$ solution above) and therefore lose one parameter.
As we will see shortly, when constructing our domain-wall geometries numerically we will choose
$V$, $W$, and $\tilde W$ to allow both modes of $\phi$ and $A_x$ in the UV.
Thus, from the UV perspective, only 1 out of the 11 integration constants
needs to be zero, and the remaining 10 parameters are free.
Armed with the behavior of the IR and UV perturbations, we are now ready to construct the interpolating domain-wall
solutions numerically.

\section{Numerics}
\label{Numerics}

Since our main interest here is in applying our construction to the quadratic band crossing model,
we will now focus on having $z=2$ in the UV and $p=2,q=3$ in the IR, although more general choices of scaling exponents
are expected to yield similar results.
To construct the domain-wall solutions we are after, we will choose the scalar potential and couplings $W$, $\tilde{W}$ to be
\begin{eqnarray}
V(\phi )&=&-10-\frac{9 \phi ^2}{8}+\frac{\phi ^6}{6144} \, , \nonumber\\
W(\phi)&=&4+\frac{19 \phi ^2}{48} \, , \qquad \tilde{W}(\phi)=\frac{\phi ^2}{2}-\frac{\phi ^4}{192} \, .
\end{eqnarray}
As can be easily checked, these expressions satisfy the various requirements we imposed in Section \ref{Setup}.
Note that we have taken the IR and UV values of the scalar field to be, respectively, $\phi_{IR}=4 \sqrt{3}$ and $\phi_{UV}=0$.
The numerical solutions we present below were found using the shooting method.
Moreover, the parameters in the model were tuned so that the UV Lifshitz fixed point would have no irrelevant modes,
to make it easier to hit by shooting from the IR.
Finally, the (irrelevant) IR modes where chosen to have the same scaling power, which we label by $\Delta$ below,
again to facilitate the numerical analysis.

For the numerical analysis we have found it convenient to use the following ansatz for the domain-wall solution,
\begin{eqnarray}\label{ansatz}
ds^2&=&-r^6 e^{f_1(r)}dt^2+\frac{dr^2}{r^2}+ r^4 e^{f_2(r)}dx^2+ r^6 e^{f_3(r)}dy^2 \, , \nonumber\\
A_t(r) &=& r^3 a_t(r) \, , \nonumber \\
A_x(r) &=&r^2 e^{\half f_2(r=0)}(1+a_x(r)) \, .
\end{eqnarray}
The IR expansion we have used to set up the numerics, which is of course based on the perturbation analysis of
Section \ref{Setup}, then takes the form
\begin{eqnarray}\label{IR expansion}
f_1(r)&= &f_{1+}+ \alpha_3 \, r ^\Delta+\ldots\nonumber\\
f_2(r)&=& f_{2+}-\frac{2}{3} \left(3+\sqrt{6}\right) \alpha_3 \, r^\Delta+\ldots\nonumber\\
f_3(r)&=& f_{3+}+ \alpha_3 \, r^\Delta+\ldots\nonumber\\
a_t(r)&=& a_{t+} \, r^\Delta + \ldots \; , \nonumber\\
a_x(r)&=& \alpha_4 \, r^\Delta-\frac{1}{3} \left(2\sqrt{6}-3\right) \alpha_3 + \ldots \; ,\nonumber\\
\phi(r)&=& 4 \sqrt{3}+\phi_+ r^\Delta + \ldots \; ,
\end{eqnarray}
where the scaling exponent is given by
$$\Delta = 2(-2 + \sqrt{6}) \, , $$
as can be read off by comparing with (\ref{IRirrelevant}).

On the other hand, the UV expansion now takes the form
\begin{eqnarray}\label{UV expansion}
f_1(r)&=& -2 \log r+f_{10}+\frac{-\frac{2\beta_4}{5}+\frac{3\tilde{\beta_4}}{5}-\frac{2\tilde{\beta_4}}{5}
\log r}{r^4} + \ldots\nonumber\\
f_2(r)&=& -2\log r +f_{20}+\frac{-\frac{2\beta_4}{5}-\beta_3-\frac{2\tilde{\beta_4}}{5}-\frac{\tilde{\beta_4}}{5}\log r}{r^4}
+ \ldots\nonumber\\
f_3(r)&=& -4\log r +f_{30}+\frac{\beta_3-\frac{\tilde{\beta_4}}{5}\log r }{r^4}+ \ldots\nonumber\\
A_t(r)&=&e^{f_{10}/2} r^2 \left(1+\frac{\beta_4+\tilde{\beta_4}\log r}{r^4} + \ldots\right)\nonumber\\
A_x(r)&=& a_{x0}+a_{x1} \, r^{-2} + \ldots\nonumber\\
\phi(r)&=& \phi_1 r^{\Delta_1}+\phi_2 r^{\Delta_2} + \ldots \; .
\end{eqnarray}
First, notice that the role of the leading log terms in the expressions for $f_1(r)$, $f_2(r)$ and $f_3(r)$
is simply to ensure that in the UV the solution is indeed the standard $z=2$ Lifshitz metric.
These terms describe the background and should not be confused with the logarithmic modes we discussed in
Section \ref{Setup}.
The remaining perturbations of the metric and of $A_t$, on the other hand, are either constant,
or suppressed by
$r^{-4}$, with the detailed structure of the latter perfectly consistent with \eqref{-4log}.
The marginally relevant log modes we discussed in Section \ref{Setup} will not be turned on in the
interpolating solution we have constructed, as will be apparent shortly.
Finally, in the expansion of the scalar field we have $\Delta_{1,2}=-2\pm\frac{\sqrt{195}}{12}$.

Notice from these expansions that we have 6 parameters describing the IR, labelled by
$\{f_{1+},f_{2+},f_{3+},\alpha_3,a_{t+},\phi_+\}$, and 10 parameter in the UV (having turned off the leading log mode)
labelled by $\{f_{10},f_{20},f_{30},\beta_3,\beta_4,\tilde\beta_4,a_{x0},a_{x1},\phi_1,\phi_2\}$.
The equation of motions in our model reduce to 1 first order ODE and 5 second order ODE,
hence we need 11 integration constants to define a solution.
We are then left with solutions parameterized by $6+10-11=5$ parameters.
Furthermore, notice that the ansatz \eqref{ansatz} has four unfixed scaling symmetries, described by:
\begin{eqnarray}
&&t\rightarrow\lambda t\text{,  }e^{f_1}\rightarrow \lambda^{-1}e^{f_1} \, , \\
&&x\rightarrow\lambda x\text{,  }e^{f_2}\rightarrow \lambda^{-1}e^{f_2} \, , \\
&&y\rightarrow\lambda y\text{,  }e^{f_3}\rightarrow \lambda^{-1}e^{f_3}\, , \\
&&r\rightarrow \lambda r\text{,  } t\rightarrow \lambda^{-3}
t\text{,  }x\rightarrow \lambda^{-2} x\text{,  }y\rightarrow \lambda^{-3} y \, .
\end{eqnarray}
After taking these into account, we are left with a \emph{one-parameter} family of solutions\footnote{In the numerical analysis we have used a different metric ansatz, which is easier to work with,
\begin{eqnarray}
ds^2&=&-r^6(1+2U(r))dt^2+\frac{dr^2}{r^2 (1+2U(r))}+ r^4 e^{2 f_2(r)}dx^2+ r^6 e^{2 f_3(r) }dy^2\nonumber
\end{eqnarray}
and have converted the result back to \eqref{ansatz} by a coordinate transformation.}.

\begin{figure}[h]
\includegraphics[width=8cm]{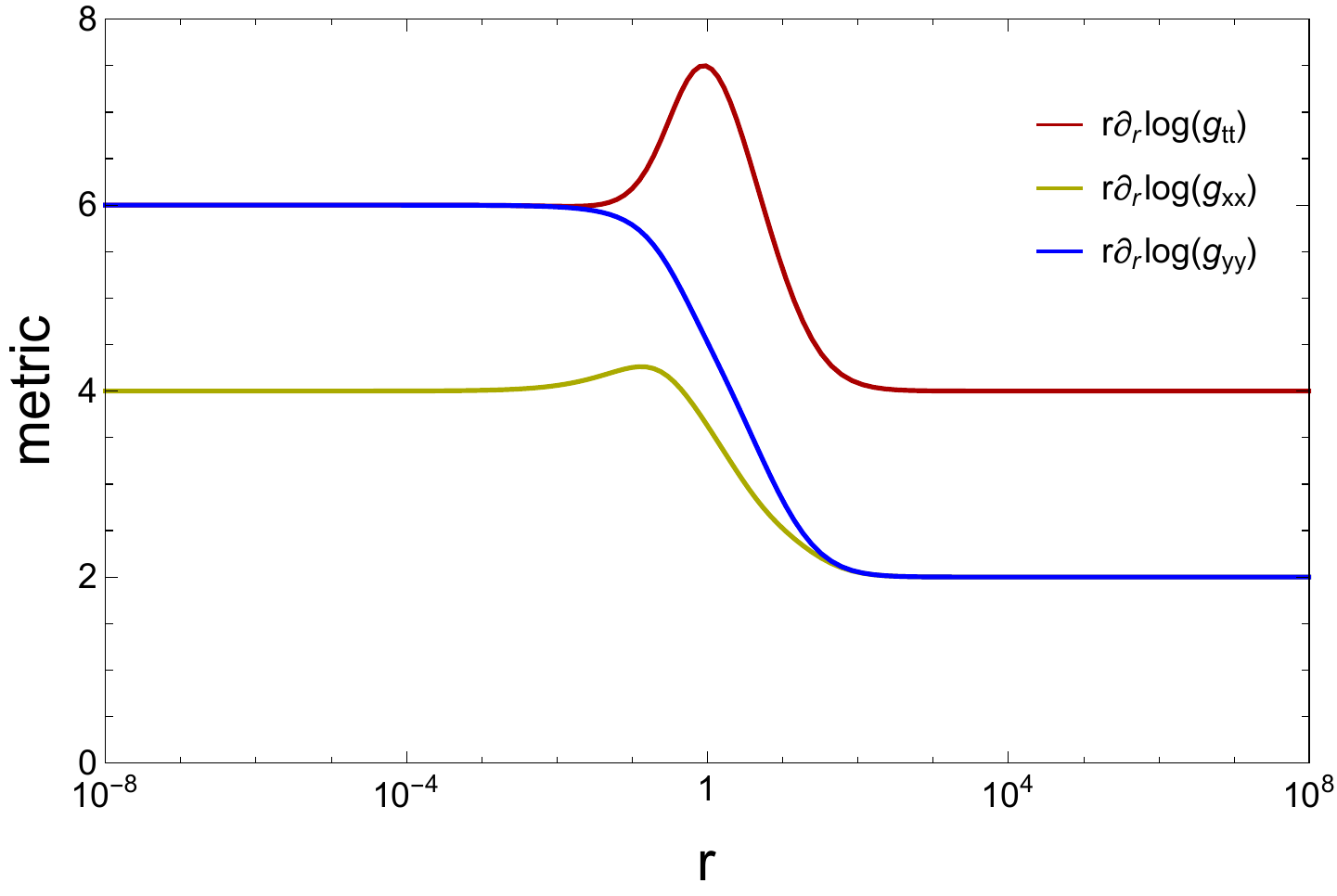}
\caption{Radial dependence of the metric components $g_{tt}, g_{xx}$ and $g_{yy}$ as they interpolate between the IR ($r=0$)
and the UV ($r\rightarrow \infty$), shown using a logarithmic derivative plot to make the IR and UV scalings transparent.
We have used a logarithmic scale on the horizontal axis.}\label{metricfigure}
\end{figure}

\begin{figure}[h]
\includegraphics[width=8cm]{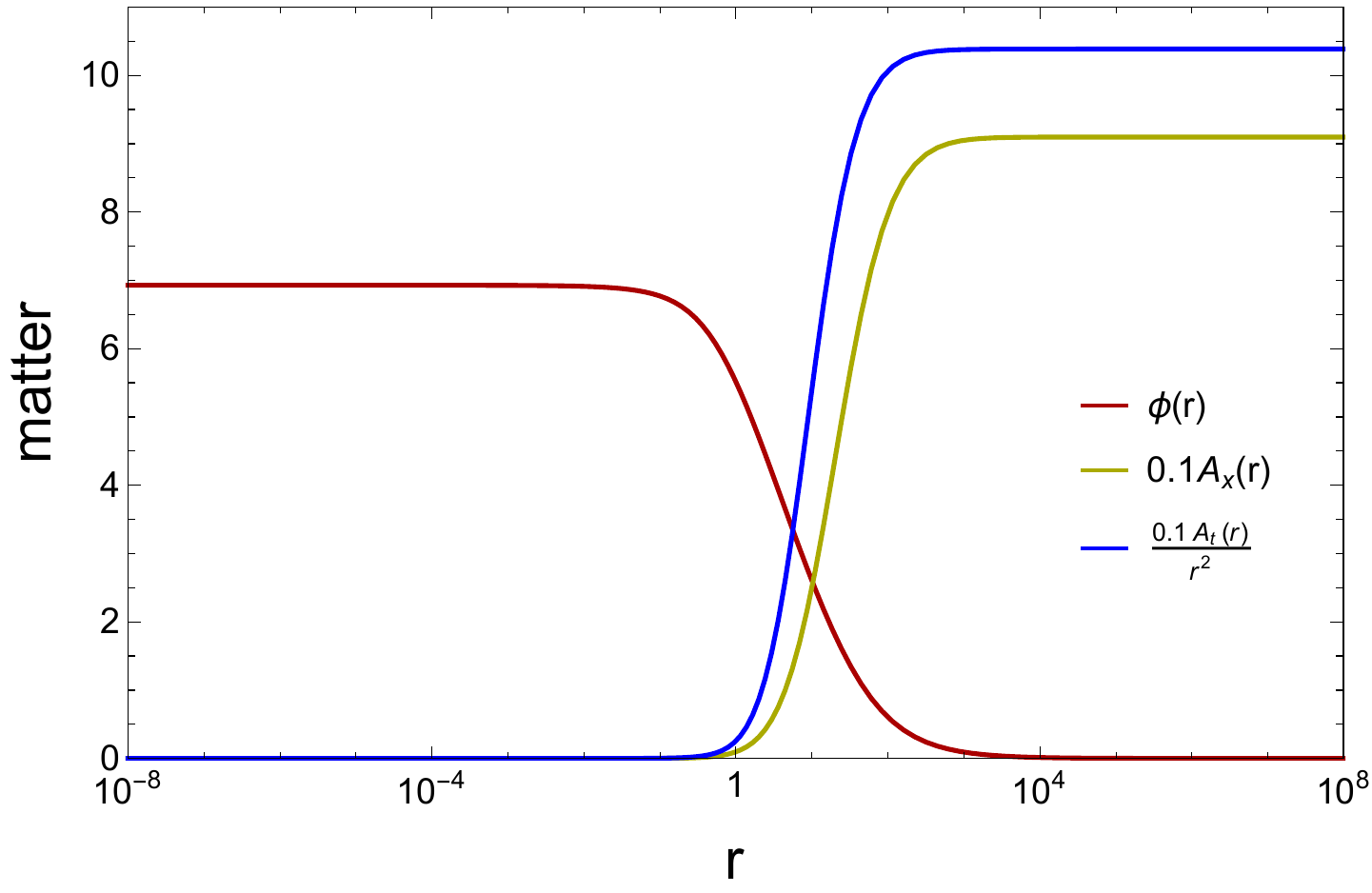}
\caption{Radial dependence of the scalar $\phi$ and gauge fields $A_t$, $A_x$ as they interpolate between the IR
($r=0$) and the UV ($r\rightarrow \infty$), shown using a logarithmic scale on the horizontal axis.
The magnitude of the gauge fields has been rescaled appropriately
in order to easily display the three matter fields in the same figure. The IR and UV scalings are consistent with
the expansions (\ref{IR expansion}) and (\ref{UV expansion}).}\label{matterfigure}
\end{figure}

In Figures \ref{metricfigure} and \ref{matterfigure} we present a typical example of the
interpolating solutions we have constructed numerically.
In these solutions the breaking of (spatial) rotational symmetry is explicit, since $a_{x0} \neq 0$.
However, in our model we should also be able to realize spontaneous symmetry breaking, since we are working with a one-parameter family of solutions.
Figure \ref{metricfigure} shows a logarithmic derivative plot\footnote{Logarithmic derivative plots are constructed
so that any function scaling as $r^\gamma$
will appear as a horizontal line with intercept at $\gamma$, making any scaling behavior readily apparent.}
of the radial dependence of the metric components $g_{tt}$, $g_{xx}$ and $g_{yy}$.
In the IR, which corresponds to $r\sim 0$,  $g_{tt}$ and $g_{yy}$ both scale as $r^6$, while $g_{xx}$ scales as
$r^4$, as expected from the nematic solution (\ref{IRbackground}) when $p=2,q=3$.
In the UV, as $r\rightarrow \infty$, it is the spatial components which scale in the same way
$g_{xx}\sim g_{yy} \sim r^2$ (showing that spatial rotations are preserved),
while $g_{tt}\sim r^4$, as expected from the Lifshitz geometry (\ref{UVbackground}) when the critical exponent is $z=2$.
Thus, we see clearly the breaking of rotational symmetry as the solution approaches the IR of the geometry.

Figure \ref{matterfigure} shows the radial dependence of the scalar field and of the two gauge fields
(appropriately rescaled).
We see that the scalar (denoted by the red line)
interpolates between $\phi_{IR} = 4\sqrt{3}$ near $r=0$ to $\phi_{UV}=0$ at the boundary, as desired.
The gauge field $A_t$ (blue line) vanishes towards the IR in agreement with the perturbation analysis (\ref{IR expansion}),
and scales as $r^2$ in the UV, as expected from (\ref{UV expansion}).
Similarly, the spatial component $A_x$ (yellow line) scales as $r^2$ in the IR and approaches a constant in the UV,
again in agreement with the IR and UV expansions (\ref{IR expansion}) and (\ref{UV expansion}).

In particular, from the UV behavior of $A_t$ it is apparent that in this background the leading log mode we discussed in Section \ref{Setup}
is not present.
Although it would be interesting to find solutions for which it is turned on, it is beyond the scope of this analysis.

\section{Conclusion}
\label{Conclusions}

Our goal in this paper was to construct a simple gravitational model which would admit zero-temperature solutions
interpolating between a UV Lifshitz fixed point and a nematic IR fixed point, in which spatial rotations are broken.
Such a setup has broad applications to non-relativistic RG flows in quantum systems with nematic IR phases.
For the specific scalings we have chosen, it can also
be thought of as a first step towards describing holographically the behavior of the bosonic modes
of the quadratic band crossing model, which
obey the Lifshitz scaling (\ref{UVscaling}) at high energies, and the nematic scaling (\ref{IRscaling})
at lower energy scales.
We emphasize that while in our numerical solutions we have chosen the scaling exponents to be $z=2$, $p=2$, and $q=3$,
our model (and the analytic results of Section \ref{Setup}) can be applied to generic values of these parameters.

The model that we have constructed couples gravity to two massive abelian gauge fields and a neutral scalar, with the latter
controlling the gauge fields' mass terms.
Our setup should by no means be the only way to engineer RG flows of this type. However,
it has the advantage of making the analysis particularly tractable, allowing us
to work with ordinary differential equations and to decouple some of the IR and UV perturbations
from the remaining ones.
While at this stage the model is entirely phenomenological, we do not see any fundamental obstacles
for being able to derive it from an appropriate supergravity truncation.

The value of the dynamical critical exponent $z=2$ we have chosen for the Lifshitz UV fixed point is interesting for several reasons.
First, it corresponds to the case of the quadratic band crossing model which is currently well-studied and understood.
Moreover, on the gravity side $z=2$ is somewhat special, since it is associated with the appearance of logarithmic
modes which have been argued to describe marginally relevant deformations of the Lifshitz theory.
Such modes affect the form of the geometry to leading order, thus modifying the Lifshitz asymptotics.
In the particular domain-wall solution we have constructed the leading log modes are turned off, and therefore do not affect
the UV behavior of the geometry. More generally, however, they are expected to be present, and it is interesting
to ask what role they may play in the physics
of the quadratic band crossing model, if any, and in nematic phases more broadly.
We leave this question to future work.

Another interesting question is that of the stability of the IR nematic geometry.
Since the spectrum of IR perturbations admits relevant modes, we expect that there should be
additional geometries that the nematic solution may flow into, perhaps
associated with the breaking of translational symmetry. Better understanding RG flow and RG stability would shed light
on the interplay between nematic and smectic phases.
Moreover, this would tie our construction to the recent efforts to probe the role of broken translations
in determining the conductive behavior of \emph{e.g.} strongly correlated electron systems.
The competition between nematic and smectic IR phases and how it connects
with the detailed behavior of the quadratic band crossing model
is an avenue that we would like to explore in future work.

\vskip 0.5in
{\noindent\large  \bf Acknowledgments}
\vskip 0.1in

We are grateful to Shamit Kachru, Jim Liu, Chris Pope and Andy Royston for helpful discussions.
We are especially thankful to Sean Hartnoll for insightful comments on the draft.
S.C. is supported by the Cambridge-Mitchell Collaboration in Theoretical Cosmology, and the Mitchell Family Foundation.
XD is supported by the National Science Foundation under grant PHY-0756174.
JR is partially supported by the Mitchell Institute for Fundamental Physics and Astronomy.
KS is supported by the National Science Foundation under grant PHY-1402971.

\end{document}